\documentclass[conference,letterpaper]{IEEEtran}
\usepackage{amsmath,amssymb,amsfonts}
\usepackage{algorithmicx}
\usepackage{algorithm,algpseudocode}
\usepackage{algcompatible}
\usepackage{array}
\usepackage{textcomp}
\usepackage{stfloats}
\usepackage{url}
\usepackage{verbatim}
\usepackage{graphicx}
\usepackage{caption}
\usepackage{cite}
\usepackage{bm}
\usepackage{xcolor}
\usepackage{colortbl}
\usepackage{multirow}
\usepackage{bbm}
\usepackage[figuresright]{rotating}
\allowdisplaybreaks

\newtheorem{example}{Example}
\newtheorem{definition}{Definition}
\newtheorem{remark}{Remark}
\ifodd 1

\else

\fi

\begin{document}
\title{Dual-Lagrange Encoding for Storage and Download in Elastic Computing for Resilience} 

\author{
\IEEEauthorblockN{Xi Zhong\textsuperscript{1},  Samuel Lu\textsuperscript{2},  Jörg Kliewer\textsuperscript{3} and Mingyue Ji\textsuperscript{1}}

\IEEEauthorblockA{\textit{\textsuperscript{1}Department of Electrical and Computer Engineering}, \textit{University of Florida}, Gainesville, FL, USA\\
Email: \{xi.zhong, mingyueji\}@ufl.edu }

\IEEEauthorblockA{\textit{\textsuperscript{2}Rowland Hall St. Marks High School}, Salt Lake City, UT, USA \\
Email: samuellu@rowlandhall.org}

\IEEEauthorblockA{\textit{\textsuperscript{3}Department of Electrical and Computer Engineering}, \textit{New Jersey Institute of Technology}, Newark, NJ, USA\\
Email: jkliewer@njit.edu }}

\maketitle

\begin{abstract}
Coded elastic computing enables virtual machines to be preempted for high-priority tasks while allowing new virtual machines to join ongoing computation seamlessly. This paper addresses coded elastic computing for matrix-matrix multiplications with straggler tolerance by encoding both storage and download using Lagrange codes.
In 2018, Yang \emph{et al.} introduced the first coded elastic computing scheme for matrix-matrix multiplications, achieving a lower computational load requirement. However, this scheme lacks straggler tolerance and suffers from high upload cost.
Zhong \emph{et al.} (2023) later tackled these shortcomings by employing uncoded storage and Lagrange-coded download. However, their approach requires each machine to store the entire dataset.
This paper introduces a new class of elastic computing schemes that utilize Lagrange codes to encode both storage and download, achieving a reduced storage size.
The proposed schemes efficiently mitigate both elasticity and straggler effects, with a storage size reduced to a fraction $\frac{1}{L}$ of Zhong \emph{et al.}'s approach, at the expense of doubling the download cost.
Moreover, we evaluate the proposed schemes on AWS EC2 by measuring computation time under two different tasks allocations: heterogeneous and cyclic assignments. Both assignments minimize computation redundancy of the system while distributing varying computation loads across machines.
\end{abstract}

\section{Introduction}
Elastic computing enables virtual machines to be preempted for high-priority tasks while incorporating new virtual machines into ongoing computations. 
An elastic computing framework typically consists of two main components: a distributed computing framework and a computation assignment. The distributed computing framework guarantees successful decoding, while the computation assignment specifies the distribution of tasks across virtual machines to minimize redundancy and accelerate computing.
Yang \textit{et al.} \cite{yang2018coded} introduced a coded elastic computing framework based on Maximum Distance Separable (MDS)-coded storage and uncoded download for matrix-vector multiplications, with cyclic computation assignments, where all machines have the same computation load and the computation redundancy is minimized. 
Following the original MDS-coded storage and uncoded download strategy, various works on the extensions such as elastic computing with heterogeneous storage or/and speeds, elastic computing tolerating stragglers, optimization on
the transition waste, were proposed in \cite{KSA2021, wcj2021hs, myjpractice, DHFL2023}.

Differing from coded storage used in \cite{yang2018coded, KSA2021, wcj2021hs, myjpractice, DHFL2023}, some works explored elastic computing with uncoded storage. 
For instance, \cite{usutec2022} introduced a framework for heterogeneous uncoded elastic computing. 
The authors in \cite{XJM2023} applied Lagrange codes to coded elastic computing for homogeneous systems under uncoded storage and coded download, which was extended to heterogeneous systems in \cite{XJM2024}. 
The authors in \cite{XJM2024} proposed a hierarchical storage placement algorithm designed to minimize expected computation time. 
More recently, \cite{HYWQJ2024} presented a decentralized elastic computing scheme with uncoded storage for heterogeneous systems. 

Despite of the advantages of uncoded download used in \cite{yang2018coded, KSA2021, wcj2021hs, myjpractice, DHFL2023, usutec2022}, it is primarily designed to address matrix-vector multiplications. Extending these approaches to matrix-matrix multiplications leads to significant download cost, as each machine must download the entire input matrix. 
In \cite{yangCEC}, a MDS-coded storage and download elastic computing scheme was proposed for matrix-matrix multiplications, but it lacks straggler tolerance and induces a high upload cost when machines send results back to the master node.
While \cite{XJM2023} addressed these limitations regarding straggler tolerance and communication cost, it requires each machine to store the entire dataset. This constraint was partially alleviated in \cite{XJM2024}, which reduces the storage size by allowing machines to store a subset of the dataset. Nonetheless, further storage reductions can be achieved through coding techniques while maintaining a low upload cost.

Motivated by challenges in straggler tolerance, computation types, storage capacity, and communication cost, we propose a new class of coded elastic computing schemes based on Lagrange-Coded Storage and Download (LCSD). Our main contributions are summarized as follows.
\begin{enumerate}
    \item Dual-Lagrange Encoding using New Partition Strategy: To address the challenges of storage size and download cost, we use Lagrange codes to encode both storage and download, i.e., encode both matrices, differing from \cite{XJM2023}, \cite{XJM2024} and \cite{XSJM2025uncodedB}.
    Ideally, encoding each matrix leads to a smaller matrix size, contributing to lower upload cost. However, in \cite{yangCEC} the decoding process fixes a large dimension of uploaded data. In this paper, our solution is to change the partition strategy of matrices so that the uploaded data has sub-matrix size, and to re-design the decoder functions ensuring successful decoding.
    \item Two LCSD Schemes: We propose two LCSD schemes which effectively mitigate the impacts of elasticity and stragglers.  Scheme $1$ focuses on reducing download cost. Scheme $2$ aims at reducing the storage size. 
    A comparison of the proposed schemes with existing schemes is summarized in Table \ref{table-complexity}.
    \item Storage-Sharing Algorithm: We integrate storage-sharing into our schemes, enabling trade-offs between storage and other metrics, such as download cost, upload cost, computing complexity and decoding complexity. Comparisons under this algorithm are also presented.
    \item AWS EC2 Experiments: Experiments on AWS EC2 show that heterogeneous assignments improve performance by $20\%$-$30\%$ over cyclic assignments in systems without straggler tolerance, and by less than $22\%$ when tolerating up to $4$ stragglers.
\end{enumerate}
 \paragraph*{Notation Convention}
$[N] = \{1, 2, \cdots, N\}$. $[a, b]$ represents the set of real numbers $c$ such that $a \leq c \leq b$.
We use $|\cdot|$ to denote the cardinality of a set.
$\mathbb{F}$ represents a finite field.

\section{System Model}
Consider a distributed system comprising a master node and a set of $N$ machines, labeled $[N]$, which collaboratively perform matrix-matrix multiplications over multiple time steps. Given a data matrix $\boldsymbol{A} \in \mathbb{F}^{q \times v}$, the task at the $t$-th time step is to compute $\boldsymbol{AB}^{(t)}$, carried out by $N_t$ available machines denoted as $\mathcal{N}_t \subseteq [N]$, where $|\mathcal{N}_t| = N_t$ and $\boldsymbol{B}^{(t)} \in \mathbb{F}^{v \times r}$. The system can tolerate up to $S$ stragglers among $N_t$ machines.
The process is as follows.
In the {\em storage placement phase}, each machine $n \in [N]$ stores a function of the data matrix $\boldsymbol{A}$.  The storage size per machine is normalized by the size of $\boldsymbol{A}$.
In each time step $t$, the master assigns computation tasks to the available machines. Each machine then downloads a function of $\boldsymbol{B}^{(t)}$ from the master. This phase is referred to as the {\em download phase}. 
The download cost per machine is defined as the size of this transmission to a machine.
In the {\em computing phase}, each machine $n \in \mathcal{N}_{t}$ processes its assigned tasks locally, and uploads the computation results back to the master. 
The upload cost per machine is defined as the size of computation results sent by a machine. 
In the {\em decoding phase}, upon receiving sufficient computation results from the available machines, the master decodes $\boldsymbol{AB}^{(t)}$ successfully,  tolerating up to $S$ stragglers without waiting for their uploads.

\setlength\belowcaptionskip{-1ex}
\begin{table*}
\centering
\begin{tabular}{|c|c|c|c|c|c|c|c|c|} 
\hline
             & Storage Size & $\mathcal{C}_{\text{Encoding}}$  & $\mathcal{C}_{\text{Download}}$ & $\mathcal{C}_{\text{Computing}}$ & $\mathcal{C}_{\text{Upload}}$ & $\mathcal{C}_{\text{Decoding}}$\\ 
\hline
Scheme $1$  & $\frac{1}{L}$ & $qv + \frac{vr(2L+S-1)}{N_t}$ & $\frac{vr(2L+S-1)}{LN_{t}}$ & $\frac{qvr(2L+S-1)}{LN_{t}}$ & $\frac{qr(2L+S-1)}{N_{t}}$ & $qrL(2L-1)$ \\
\hline
Scheme $2$ & $\frac{2L+S-1}{LN_{t}}$ & $\frac{qv(2L+S-1)}{N_{t}} + vr$ & $\frac{vr}{L}$ & $\frac{qvr(2L+S-1)}{LN_{t}}$ & $\frac{qr(2L+S-1)}{N_{t}}$  & $qrL(2L-1)$ \\
\hline
\hline
 \cite{yang2018coded}  & $\frac{1}{L}$ & $qv$ &  $vr$ & $\frac{qvr(L+S)}{LN_{t}}$  & $\frac{qr(L+S)}{LN_{t}}$  &  $qrL$\\
\hline
\cite{XJM2023}   & $1$  & $\frac{vr(L+S)}{N_{t}}$ & $\frac{vr(L+S)}{LN_{t}}$ & $\frac{qvr(L+S)}{LN_{t}}$ &  $\frac{qr(L+S)}{LN_{t}}$ &  $qrL$\\
\hline
 \cite{XJM2024}   & $\frac{L+S}{N_{t}}$  & $vr$ & $\frac{vr}{L}$ & $\frac{qvr(L+S)}{LN_{t}}$ &  $\frac{qr(L+S)}{LN_{t}}$ &  $qrL$\\
 \hline
\cite{yangCEC}   & $\frac{1}{L}$ & $qv + \frac{vrL}{N_{t}}$ &  $\frac{vr}{N_{t}}$ & $\frac{qvr}{N_{t}}$ & $qrL$ & $\mathcal{O}(1)$\\
\hline
\end{tabular}
\caption{Using cyclic assignment, we compare the LCSD schemes with several existing schemes, in terms of $6$ metrics, including storage size per machine, encoding complexity at the master for each machine (denoted as $\mathcal{C}_\text{Encoding}$), download cost per machine (denoted as $\mathcal{C}_{\text{Download}}$), computing complexity per machine (denoted as $\mathcal{C}_{\text{Computing}}$), upload cost per machine (denoted as $\mathcal{C}_{\text{Upload}}$), and decoding complexity at the master (denoted as $\mathcal{C}_{\text{Decoding}}$).
\label{table-complexity} }
\end{table*}

\section{Proposed LCSD Systems}
\label{sec: fix}
We present LCSD systems in two steps. 
1) We first address the case where the set of available machines, $\mathcal{N}_t$, remains fixed in all time steps. The key differences between the two proposed schemes are described, followed by a detailed description of the general schemes.
2) we consider a scenario where the system can tolerate up to $P$ unavailable machines due to elasticity.

We consider a system with a fixed $\mathcal{N}_{t}$ for any time step $t$. 
We select $L$ numbers $\{ \beta_{l} \in \mathbb{F}:$ $l \in [L]\}$ with $2L+S-1 \leq N_{t}$, and $N$ numbers $\{\alpha_{n} \in \mathbb{F}$ $ : n \in [N]\}$, such that $\{\alpha_{n} :$ $n \in [N]\} \cap$ $\{ \beta_{l} \in \mathbb{F}: l \in [L]\} = \emptyset$. Assign each machine $n \in [N]$ to a unique $\alpha_n$. 
We divide the data matrix $\boldsymbol{A}$ column-wise and matrix $\boldsymbol{B}^{(t)}$ row-wise into $L$ equal-sized sub-matrices, respectively, denoted by
\begin{equation}
\begin{split}
    \label{eq-partition-L}
    & \boldsymbol{A}  = [\boldsymbol{A}_{1}, \boldsymbol{A}_{2}, \cdots, \boldsymbol{A}_{L}], \\
    & \boldsymbol{B}^{(t)} = [(\boldsymbol{B}^{(t)}_{1})^T, (\boldsymbol{B}^{(t)}_{2})^T, \cdots, (\boldsymbol{B}^{(t)}_{L})^T]^T
\end{split}
\end{equation}
To explain the application of Lagrange codes and highlight the differences between the two proposed schemes, we introduce traditional Lagrange-coded computing as a foundation.

{\bf Traditional Lagrange-Coded Computing: } 
We consider the following polynomials, each of degree $L-1$,
\begin{align}
    V(z) & = \sum_{l \in [L]} \boldsymbol{A}_{l} \cdot \prod_{l'\in [L] \setminus {\{l\}}} {\frac{z-\beta_{l'}}{\beta_{l}-\beta_{l'}}}, \label{eq-ex-encodingA} \\
    U(z) & = \sum_{l \in [L]} \boldsymbol{B}^{(t)}_{l} \cdot \prod_{l'\in [L] \setminus {\{l\}}} {\frac{z-\beta_{l'}}{\beta_{l}-\beta_{l'}}}, \label{eq-ex-encodingB}
\end{align}
which satisfy $V(\beta_{l}) = \boldsymbol{A}_{l}$ and  $U(\beta_{l}) = \boldsymbol{B}^{(t)}_{l}$ for $l \in [L]$. Recall that $\boldsymbol{AB}^{(t)}$ $ =$ $ \sum_{l \in [L]}$ $\boldsymbol{A}_{l}\boldsymbol{B}^{(t)}_{l}$. Thus, $\boldsymbol{AB}^{(t)}$ $=$ $\sum_{l \in [L]}$ $V(\beta_{l})U(\beta_{l})$. 
To recover $V(\beta_{l})U(\beta_{l})$ for $l \in [L]$, the coded computing scheme can be designed as follows.
Let machine $n \in \mathcal{N}_{t}$ store the coded matrix $V(\alpha_{n})$ during the storage placement phase, and receive the coded matrix $U(\alpha_{n})$ during the download phase, where we denote $V(\alpha_{n}) = \Tilde{\boldsymbol{A}}_{n}$ and $U(\alpha_{n}) = \Tilde{\boldsymbol{B}}_{n}$. In the computing phase, machine $n \in \mathcal{N}_{t}$ computes $\Tilde{\boldsymbol{A}}_{n}\Tilde{\boldsymbol{B}}_{n}$.
In the decoding phase, we consider the polynomial $V(z)U(z)$ of degree $2L-2$. Each $V(\beta_{l})U(\beta_{l})$ for $l \in [L]$ is an evaluation of $V(z)U(z)$, and each computation result $ \Tilde{\boldsymbol{A}}_{n}\Tilde{\boldsymbol{B}}_{n}$ for $n \in \mathcal{N}_{t}$ is an evaluation of $V(z)U(z)$ due to $\Tilde{\boldsymbol{A}}_{n}\Tilde{\boldsymbol{B}}_{n} = V(\alpha_n)U(\alpha_n)$. 
Hence, $V(\beta_{l})U(\beta_{l})$ can be decoded, by interpolating the polynomial $V(z)U(z)$ using any $2L-1$ computation results and evaluating it on $\beta_{l}$. Finally, $\boldsymbol{AB}^{(t)}$ $=$ $\sum_{l \in [L]} V(\beta_{l})U(\beta_{l})$ is decoded.

However, this scheme requires machines to return at least $N_{t} - S$ computation results, while only $2L-1 \leq N_{t} -S$ results are sufficient for successful decoding. To address this redundancy, we propose two new schemes that minimize computation redundancy by allowing each machine to compute a subset of the computation task $\Tilde{\boldsymbol{A}}_{n}\Tilde{\boldsymbol{B}}_{n}$, which correspondingly reduces both the storage size and the download cost.
The main difference between the two schemes is the strategies for splitting $\Tilde{\boldsymbol{A}}_{n}\Tilde{\boldsymbol{B}}_{n}$ into sub-tasks. 

Specifically, the computation $\Tilde{\boldsymbol{A}}_{n}\Tilde{\boldsymbol{B}}_{n}$ can be  divided into $G$ sub-tasks using three distinct partitioning strategies. 
Partitioning Strategy $1$: The download $\Tilde{\boldsymbol{B}}_{n}$ is divided column-wise into $G$ sub-matrices, denoted by $\Tilde{\boldsymbol{B}}_{n} = $ $[\Tilde{\boldsymbol{B}}_{n, 1}$,   $\Tilde{\boldsymbol{B}}_{n, 2} $,  $\cdots $,  $\Tilde{\boldsymbol{B}}_{n, G}]$. Thus, $\Tilde{\boldsymbol{A}}_{n}\Tilde{\boldsymbol{B}}_{n} = $  $[\Tilde{\boldsymbol{A}}_{n}\Tilde{\boldsymbol{B}}_{n, 1} $,   $\Tilde{\boldsymbol{A}}_{n}\Tilde{\boldsymbol{B}}_{n, 2} $,  $\cdots $,  $\Tilde{\boldsymbol{A}}_{n}\Tilde{\boldsymbol{B}}_{n, G}]$.
In proposed Scheme $1$, each machine computes sub-tasks $\Tilde{\boldsymbol{A}}_{n}\Tilde{\boldsymbol{B}}_{n, g}$ for some $g \in [G]$. 
Partitioning Strategy $2$: The storage $\Tilde{\boldsymbol{A}}_{n}$ is row-wise divided to $G$ sub-matrices, denoted by $\Tilde{\boldsymbol{A}}_{n} = $ $[(\Tilde{\boldsymbol{A}}_{n, 1})^T$,   $(\Tilde{\boldsymbol{A}}_{n, 2})^T $,  $\cdots $,  $(\Tilde{\boldsymbol{A}}_{n, G})^T]^T$. Thus, $\Tilde{\boldsymbol{A}}_{n}\Tilde{\boldsymbol{B}}_{n} = $  $[(\Tilde{\boldsymbol{A}}_{n,1}\Tilde{\boldsymbol{B}}_{n})^T $,   $(\Tilde{\boldsymbol{A}}_{n,2}\Tilde{\boldsymbol{B}}_{n})^T$,  $\cdots $,  $(\Tilde{\boldsymbol{A}}_{n, G}\Tilde{\boldsymbol{B}}_{n})^T]^T$. In proposed Scheme $2$, each machine computes sub-tasks $\Tilde{\boldsymbol{A}}_{n,g}\Tilde{\boldsymbol{B}}_{n}$ for some $g \in [G]$.
Partitioning Strategy $3$: The storage $\Tilde{\boldsymbol{A}}_{n}$ is column-wise divided into $G$ sub-matrices, i.e.,$\Tilde{\boldsymbol{A}}_{n} = $ $[\Tilde{\boldsymbol{A}}_{n, 1}$,   $\Tilde{\boldsymbol{A}}_{n, 2} $,  $\cdots $,  $\Tilde{\boldsymbol{A}}_{n, G}]$. The download $\Tilde{\boldsymbol{B}}_{n}$ is correspondingly row-wise divided into $G$ sub-matrices, i.e.,   
 $\Tilde{\boldsymbol{B}}_{n} = $ $[(\Tilde{\boldsymbol{B}}_{n, 1})^T$,   $(\Tilde{\boldsymbol{B}}_{n, 2})^T$,  $\cdots $,  $(\Tilde{\boldsymbol{B}}_{n, G})^T]^T$. Thus, $\Tilde{\boldsymbol{A}}_{n}\Tilde{\boldsymbol{B}}_{n} = \sum_{g\in [G]}\Tilde{\boldsymbol{A}}_{n,g}\Tilde{\boldsymbol{B}}_{n,g}$.

Using a partitioning strategy changes the storage placement and download, impacting storage size, communication costs, and computational complexity. For example, using Partitioning Strategy $1$, the download cost per machine is reduced, as machine $n \in \mathcal{N}_{t}$ receives only a subset of $\Tilde{\boldsymbol{B}}_{n}$. The storage size per machine remains $\frac{1}{L}$, as each machine $n$ stores $\Tilde{\boldsymbol{A}}_{n}$.
Using Partitioning Strategy $2$, the storage size per machine is reduced, as each machine stores only a subset of $\Tilde{\boldsymbol{A}}_{n}$.
Using Partitioning Strategy $3$, both the storage size and download cost are reduced, while at the expense of a significant increase in the upload cost per machine. Each computation result, $\Tilde{\boldsymbol{A}}_{n,g}\Tilde{\boldsymbol{B}}_{n,g}$, has a size of $qr$, equivalent to the size of $\boldsymbol{AB}^{(t)}$. Since each machine uploads multiple computation results to the master, the overall upload cost becomes substantially higher. The matrix-matrix multiplications scheme proposed in \cite{yangCEC} incurs a large upload cost because it utilizes Partitioning Strategy $3$ for dividing computation tasks. Therefore, in this paper, we focus on Partitioning Strategies $1$ and $2$.

Before presenting the general LCSD schemes,  we introduce the definition of computation assignment, which will be used to specify the sub-tasks assigned to  machines.
\begin{definition}
    \label{def-assignment}
      $(\boldsymbol{\gamma}, \boldsymbol{\mathcal{M}})$ is the computation assignment of $\mathcal{N}_{t}$, where $\boldsymbol{\gamma} = (\gamma_1, \gamma_2, \cdots, \gamma_G)$, $0 \leq \gamma_g \leq 1$ for $g \in [G]$ and $\sum_{g \in [G]} \gamma_g = 1$. $\boldsymbol{\mathcal{M}} = \{ \mathcal{M}_{1}, \mathcal{M}_{2}, \cdots, \mathcal{M}_{G}\} $, where $\mathcal{M}_{g} \subseteq \mathcal{N}_{t}$ and $\mathcal{M}_{g} = |2L+S-1|$ for $g \in [G]$. We define $\mathcal{L}_{g}$ as any subset of $\mathcal{M}_{g}$ with $|\mathcal{L}_{g}| = 2L-1$.
\end{definition}

\begin{remark}
\label{re-cyclic}
(Cyclic Assignment \cite{yang2018coded})
    $G = N_{t}$. $\boldsymbol{\gamma} = (\frac{1}{N_t}$, $\frac{1}{N_t}$, $\cdots$, $\frac{1}{N_t})$. $\mathcal{M}_{g} = \{n_{g \% N_{t}}$, $n_{(g+1) \% N_{t}}$, $\cdots$, $n_{(g+2L+S-2) \% N_{t}}\}$ for $g \in [G]$, where $n_{i}$ is the $i$-th machine in $\mathcal{N}_{t}$ and we define $a \% N_{t} = a - \lfloor \frac{a-1}{N_{t}}\rfloor N_{t} $.
\end{remark}
\begin{remark}
\label{re-assi}
    (Heterogeneous Assignment \cite{wcj2021hs}) 
    When machines have different computation speeds, $(\boldsymbol{\gamma}, \boldsymbol{\mathcal{M}})$ is obtained using Algorithm $1$ in \cite{wcj2021hs}. Specifically, given the output of Algorithm $1$, i.e., $F$, $\{\alpha_{1}, \alpha_{2}, \cdots, \alpha_{F}\}$ and $\{\mathcal{P}_{1}, \mathcal{P}_{2}, \cdots, \mathcal{P}_{F}\}$,   we let $G = F$, $\gamma_{g} = \alpha_{g}$ and $\mathcal{M}_{g} = \mathcal{P}_{g}$ for $g \in [G]$. 
\end{remark}

\subsection{LCSD Scheme $1$}
LCSD Scheme $1$ is derived from Partitioning Strategy $1$, designed to reduce download cost. Next, we redesign the encoder functions, computation tasks, and decoder functions.
\subsubsection{Storage Placement Phase} 
Machine $n \in \mathcal{N}_{t}$ stores $V(\alpha_{n}) = \Tilde{\boldsymbol{A}}_{n}$, where $V(z)$ is as defined in \eqref{eq-ex-encodingA}. 

\subsubsection{Download Phase} 
Given $(\boldsymbol{\gamma}, \boldsymbol{\mathcal{M}})$, we partition each $\boldsymbol{B}^{(t)}_{l}$ for $l \in [L]$ in \eqref{eq-partition-L} column-wise into $G$ sub-matrices based on $\boldsymbol{\gamma}$,  denoted by $\boldsymbol{B}^{(t)}_{l} =$ $[ \boldsymbol{B}^{(t)}_{l,1}$, $\boldsymbol{B}^{(t)}_{l,2}$, $\cdots$, $\boldsymbol{B}^{(t)}_{l,G}]$, where $\boldsymbol{B}^{(t)}_{l,g}$ has dimension $\frac{v}{L} \times r\gamma_g$ for $g \in [G]$.
We consider the following $G$ polynomials, each of degree $L-1$,
\begin{equation}
\label{eq-ex1-encodingB}
    U_{g}(z) = \sum_{l \in [L]} \boldsymbol{B}^{(t)}_{l, g} \cdot \prod_{l'\in [L] \setminus {\{l\}}} {\frac{z-\beta_{l'}}{\beta_{l}-\beta_{l'}}}, \text{ for } g \in [G], 
\end{equation}
which satisfies $U_{g}(\beta_{l}) = \boldsymbol{B}^{(t)}_{l,g}$ for $l \in [L]$. 
Machine $n \in \mathcal{N}_{t}$ will download evaluations $U_{g}(\alpha_{n})$ for some $g \in [G]$.
Specifically,  based on $\boldsymbol{\mathcal{M}}$ each machine $n \in \mathcal{N}_{t}$ downloads $\{U_{g}(\alpha_{n}): n \in \mathcal{M}_{g}, g \in [G]\}$. 
We denote $U_{g}(\alpha_{n}) = \Tilde{\boldsymbol{B}}_{n, g}$.
\subsubsection{Computing Phase} 
Each machine $n \in \mathcal{N}_{t}$ computes $\{\Tilde{\boldsymbol{A}}_{n} \Tilde{\boldsymbol{B}}_{n,g}$ $: n \in \mathcal{M}_{g}, g \in [G]\}$, and uploads the computation results back to the master.
\subsubsection{Decoding Phase}
Recall that $\boldsymbol{AB}^{(t)} =$ $ \sum_{l \in [L]} \boldsymbol{A}_{l} \boldsymbol{B}^{(t)}_{l}$ $=$ $\sum_{l \in [L]} \boldsymbol{A}_{l}$ $[\boldsymbol{B}^{(t)}_{l,1}$, $\boldsymbol{B}^{(t)}_{2}$, $\cdots$, $\boldsymbol{B}^{(t)}_{l,G}] =$ $[\sum_{l \in [L]}\boldsymbol{A}_{l} \boldsymbol{B}^{(t)}_{l,1}$,  $\sum_{l \in [L]}\boldsymbol{A}_{l} \boldsymbol{B}^{(t)}_{l,2}$, $\cdots$, $\sum_{l \in [L]}\boldsymbol{A}_{l} \boldsymbol{B}^{(t)}_{l,G}]$.
Next, the master recovers the block $\sum_{l \in [L]} \boldsymbol{A}_{l} \boldsymbol{B}^{(t)}_{l,g}$ using the computation results from machines $\mathcal{M}_{g}$ for each $g \in [G]$. We define the following $G$ polynomials, each of degree $2L-2$,
\begin{equation}
\label{eq-ex1-h}
    V(z)U_{g}(z), \text{ for } g \in [G].
\end{equation}
For $g \in [G]$ and $l \in [L]$, we have $V(\beta_{l})U_{g}(\beta_{l}) = \boldsymbol{A}_{l}\boldsymbol{B}^{(t)}_{l,g}$ from \eqref{eq-ex-encodingA} and \eqref{eq-ex1-encodingB}. That is, the block $\boldsymbol{A}_{l}\boldsymbol{B}^{(t)}_{l,g}$ is an evaluation of the polynomial $V(z)U_{g}(z)$. In addition, the computation results from machines $\mathcal{M}_{g}$ are evaluations of $V(z)U_{g}(z)$, as $V(\alpha_{n})U_{g}(\alpha_{n})$ $=$ $\Tilde{\boldsymbol{A}}_{n}\Tilde{\boldsymbol{B}}_{n,g}$ for $n \in \mathcal{M}_{g}$. 
Hence, decoding $\boldsymbol{A}_{l}\boldsymbol{B}^{(t)}_{l,g}$ is to evaluate $V(\beta_{l})U_{g}(\beta_{l})$ using any $2L-1$ out of $2L+S-1$ computation results from machines $\mathcal{M}_{g}$.
Using Lagrange interpolation, the master computes $\sum_{l \in [L]} \left(\sum_{n\in \mathcal{L}_{g}} \Tilde{\boldsymbol{A}}_{n}\Tilde{\boldsymbol{B}}_{n,g} \cdot \prod_{n' \in \mathcal{L}_{g} \setminus \{n\}} \frac{\beta_{l}- \alpha_{n'}}{\alpha_{n}-\alpha_{n'}}  \right) =V(\beta_{l})U_{g}(\beta_{l}) = \sum_{l \in [L]} \boldsymbol{A}_{l}\boldsymbol{B}^{(t)}_{l,g} $.
By obtaining $\sum_{l \in [L]} \boldsymbol{A}_{l}\boldsymbol{B}^{(t)}_{l,g}$ for all $g \in [G]$, $\boldsymbol{A}\boldsymbol{B}^{(t)}$ is decoded successfully.

\subsection{LCSD Scheme $2$}
LCSD Scheme $2$ is derived from Partitioning Strategy $2$, designed to reduce the storage size. Next, we redesign the encoder functions, computation tasks, and decoder functions.
\subsubsection{Storage Placement Phase}
Given $(\boldsymbol{\gamma}, \boldsymbol{\mathcal{M}})$,  we partition each $\boldsymbol{A}_{l}$ for $l \in [L]$ in \eqref{eq-partition-L} row-wise into $G$ sub-matrices based on $\boldsymbol{\gamma}$, denoted by $\boldsymbol{A}_{l} = [\boldsymbol{A}^T_{l,1}, \boldsymbol{A}^T_{l,2}, \cdots, \boldsymbol{A}^T_{l,G}]^T$, where $\boldsymbol{A}_{l,g}$ has dimensions $q\gamma_g \times \frac{v}{L}$ for $g \in [G]$. We consider the following $G$ polynomials, each of degree $L-1$,
\begin{equation}
    \label{eq-ex2-encodingA}
    V_{g}(z) = \sum_{l \in [L]} \boldsymbol{A}_{l, g} \cdot \prod_{l'\in [L] \setminus {\{l\}}} {\frac{z-\beta_{l'}}{\beta_{l}-\beta_{l'}}}, \text{ for } g \in [G], 
\end{equation}
which satisfies $V_{g}(\beta_{l}) = \boldsymbol{A}_{l,g}$ for $l \in [L]$. Based on $\boldsymbol{\mathcal{M}}$, machine $n \in \mathcal{N}_{t}$ stores evaluations $\{V_{g}(\alpha_{n}): n \in \mathcal{M}_{g}, g \in [G]\}$. We denote $V_{g}(\alpha_{n}) = \Tilde{\boldsymbol{A}}_{n, g}$.

\subsubsection{Download Phase} 
Each machine $n \in \mathcal{N}_{t}$ downloads $U(\alpha_{n}) = \Tilde{\boldsymbol{B}}_{n}$, where $U(z)$ is defined as \eqref{eq-ex-encodingB}.

\subsubsection{Computing Phase}
Each machine $n \in \mathcal{N}_{t}$ computes $\{\Tilde{\boldsymbol{A}}_{n, g} \Tilde{\boldsymbol{B}}_{n}: n \in \mathcal{M}_{g}, g \in [G]\}$, and uploads the computation results back to the master.

\subsubsection{Decoding Phase}
Recall that
$\boldsymbol{AB}^{(t)}  = \sum_{l \in [L]} \boldsymbol{A}_{l}\boldsymbol{B}^{(t)}_{l}$ $= 
        \sum_{l \in [L]} 
        \begin{bmatrix}
            \boldsymbol{A}_{l,1} \\
            \boldsymbol{A}_{l,2} \\
            \vdots \\
            \boldsymbol{A}_{l,G} \\
        \end{bmatrix} \boldsymbol{B}_{l} = 
        \begin{bmatrix}
            \sum_{l\in [L]} \boldsymbol{A}_{l,1}\boldsymbol{B}_{l} \\
            \sum_{l\in [L]} \boldsymbol{A}_{l,2}\boldsymbol{B}_{l} \\
            \vdots \\
            \sum_{l\in [L]} \boldsymbol{A}_{l,G}\boldsymbol{B}_{l} 
        \end{bmatrix}$.
Next, the master recovers the block $\sum_{l \in [L]} \boldsymbol{A}_{l,g}\boldsymbol{B}_{l}$ using the computation results from machines $\mathcal{M}_{g}$ for each $g \in [G]$. Specifically, we define the following $G$ polynomials, each of degree $2L-2$,
\begin{equation}
    \label{eq-ex2-h}
    V_{g}(z) U(z), \text{ for } g \in [G].
\end{equation}
For $g \in [G]$ and $l \in [L]$, we have $V_{g}(\beta_{l})U(\beta_{l}) = \boldsymbol{A}_{l,g}\boldsymbol{B}^{(t)}_{l}$ from \eqref{eq-ex-encodingB} and \eqref{eq-ex2-encodingA}. That is, the block $\boldsymbol{A}_{l,g}\boldsymbol{B}^{(t)}_{l}$ is an evaluation of the polynomial $V_{g}(z) U(z)$. 
In addition, the computation results from machines $\mathcal{M}_{g}$ are evaluations of  $V_{g}(z) U(z)$, as $V_{g}(\alpha_{n})U(\alpha_{n}) =  \Tilde{\boldsymbol{A}}_{n,g}\Tilde{\boldsymbol{B}}_{n}$ for $n \in \mathcal{M}_{g}$. 
Hence, using Lagrange interpolation, the master computes $\sum_{l \in [L]} \left(\sum_{n\in \mathcal{L}_{g}} \Tilde{\boldsymbol{A}}_{n, g}\Tilde{\boldsymbol{B}}_{n} \cdot \prod_{n' \in \mathcal{L}_{g} \setminus \{n\}} \frac{\beta_{l}- \alpha_{n'}}{\alpha_{n}-\alpha_{n'}}  \right)$ $=$ $V_{g}(\beta_{l})U(\beta_{l})$ $=$ $\sum_{l \in [L]} \boldsymbol{A}_{l,g}\boldsymbol{B}^{(t)}_{l} $. By obtaining $\sum_{l \in [L]}\boldsymbol{A}_{l,g}\boldsymbol{B}_{l}$ for all $g \in [G]$, $\boldsymbol{A}\boldsymbol{B}^{(t)}$ is decoded successfully.

In both Scheme $1$ and Scheme $2$, recovering $\boldsymbol{AB}^{(t)}$ requires decoding the $G$ blocks contained within $\boldsymbol{AB}^{(t)}$. Each block is decoded by evaluating the points $\{\beta_{l}: l \in [L]\}$ on the polynomials in \eqref{eq-ex1-h} and \eqref{eq-ex2-h}, respectively.  $|\mathcal{L}_{g}| = 2L-1$ computation results are sufficient for the master to successfully decode each block in $\boldsymbol{AB}^{(t)}$, as the degrees of the polynomials in \eqref{eq-ex1-h} and \eqref{eq-ex2-h} are $2L-2$. The design ensures that $|\mathcal{M}_{g}| - |\mathcal{L}_{g}| = S$, enabling the system to tolerate up to $S$ stragglers.

Next, we extend the proposed schemes to the scenario where the system tolerates up to $P$ unavailable machines.
\subsection{Storage Placement for Tolerating $P$ Unavailable Machines}
In the proposed LCSD schemes, the recovery threshold, i.e., the minium number of machines required for successful decoding, is $2L-1$. Therefore, for successful decoding and straggler tolerance of $S$, $N_{t} \geq 2L+S-1$ must hold for any time step $t$.
We denote $P$ as the maximum number of preempted machines the system can tolerate, meaning $N_{t} \geq N - P$ for any time step $t$. 
Thus, $P$ can range from $0$ to $N-(2L+S-1)$, i.e.,  $P \in \{0, 1, \cdots, N-(2L+S-1)\}$. 
The goal is to determine the storage placement of the system, supporting it tolerates up to any $P$ unavailable machines.

Given $P$, the set of all available realizations is $\boldsymbol{\mathcal{N}}_{P} = \{\mathcal{N} :  \mathcal{N} \subseteq [N],  N - P \leq  |\mathcal{N}| \leq N \}$, meaning that $\mathcal{N}_{t} \in \boldsymbol{\mathcal{N}}_{P}$ for any time step $t$. The size of $\boldsymbol{\mathcal{N}}_{P}$ is given by $|\boldsymbol{\mathcal{N}}_{P}|  = {N \choose 0} + {N \choose 1} + \cdots + {N \choose P}$. 
The storage placement is designed as follows. 
Each machine applies the union of its storage placements across all availability realizations in $\boldsymbol{\mathcal{N}}_{P}$.
Specifically, the storage of machine $n$ is defined as $\bigcup_{i \in [\left|\boldsymbol{\mathcal{N}}_{P}\right|]} \mathcal{S}_{n,i}$, where $\mathcal{S}_{n,i}$ denotes the storage placement of machine $n$ for the $i$-th availability realization in $\boldsymbol{\mathcal{N}}_{P}$.

\section{Simulations and Experiments on AWS EC2}
\subsection{Computational Complexity based on Storage-sharing}
\label{sec-storage-sharing}
From Table \ref{table-complexity}, the proposed schemes are limited to specific storage sizes because $L$ must be an integer. To address this limitation, we introduce a storage-sharing approach that enables flexible storage sizes. This method is outlined in Algorithm \ref{al-sharing}, where $\boldsymbol{A}_{\lambda}$ has $\lambda q$ rows and $\boldsymbol{A}_{1-\lambda}$ has $(1-\lambda)q$ rows in lines $4$ and $11$. 
\begin{algorithm}
  \caption{Storage-Sharing of Scheme $i$ and Scheme $j$}
  \label{al-sharing}
  \begin{algorithmic}[1]
  \item [{\bf Input}: Scheme $i$, Scheme $j$, $L'$]
   \hspace*{4cm} 
   \IF{Scheme $i =$ Scheme $j$}
     \FOR{$L = L', L'-1, ..., 3$}
         \FOR{$\lambda : 1 \rightarrow 0$}
            \STATE $\boldsymbol{A} = \begin{bmatrix} \boldsymbol{A}_{\lambda} \\ \boldsymbol{A}_{1- \lambda} \end{bmatrix}$
            \STATE Use Scheme $i$ for $\boldsymbol{A}_{\lambda}\boldsymbol{B}^{(t)}$ with parameter $L$
            \STATE Use Scheme $j$ for $\boldsymbol{A}_{1-\lambda}\boldsymbol{B}^{(t)}$ with parameter $L-1$
        \ENDFOR
    \ENDFOR
    \ELSE
        \FOR{$\lambda : 1 \rightarrow 0$}
            \STATE $\boldsymbol{A} = \begin{bmatrix} \boldsymbol{A}_{\lambda} \\ \boldsymbol{A}_{1- \lambda} \end{bmatrix}$
            \STATE Use Scheme $i$ for $\boldsymbol{A}_{\lambda}\boldsymbol{B}^{(t)}$ with parameter $L'$
            \STATE Use Scheme $j$ for $\boldsymbol{A}_{1-\lambda}\boldsymbol{B}^{(t)}$ with parameter $L'$
        \ENDFOR
    \ENDIF
\end{algorithmic}
\end{algorithm}
For example, when $i = j = 1$,  the storage size of Scheme $1$ can be adjusted within the range $\left[\frac{1}{L'}, \frac{1}{2}\right]$ for an integer $L'$. When $i = 1$ and $j = 2$, the storage-sharing between Scheme $1$ and Scheme $2$ achieves a storage size within  $\left[ \frac{2L'+S-1}{L'N_{t}}, \frac{1}{L'}\right]$ given integers $L'$, $S$ and $N_{t}$. 
\begin{example}
\label{ex-storage-sharing}
    When $q = v = r = 500$, $L' = 9$, $\mathcal{N}_{t} = [21]$ and $i = j = 1$, using Algorithm \ref{al-sharing} the storage size of Scheme $1$ is within the range $\left[\frac{1}{9}, \frac{1}{2} \right]$. The resulting trade-offs between storage size per machine and other performance metrics are illustrated in Fig. \ref{fig-sharing-1}.
    \setlength\belowcaptionskip{-2ex}
    \begin{figure}
    \centerline{\includegraphics[scale= 0.6]{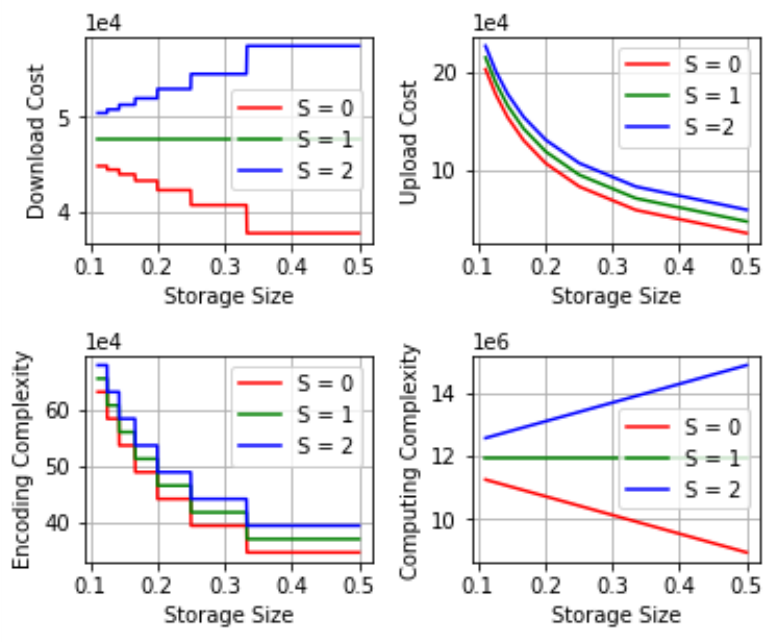}}
    \caption{Storage-sharing of Scheme $1$ in Example \ref{ex-storage-sharing}. The red, green, and blue lines represent to the cases for $S = 0$, $S =1$ and $S=2$, respectively. }
    \label{fig-sharing-1}
    \end{figure}
    For comparison, when $S = 0$, we apply the schemes in \cite{yang2018coded} and \cite{yangCEC} to storage-sharing, by executing lines $1$-$8$ for the schemes in \cite{yang2018coded} and \cite{yangCEC}, respectively, as depicted in Fig. \ref{fig-sharing-compare}.
     \begin{figure}
    \centerline{\includegraphics[scale= 0.6]{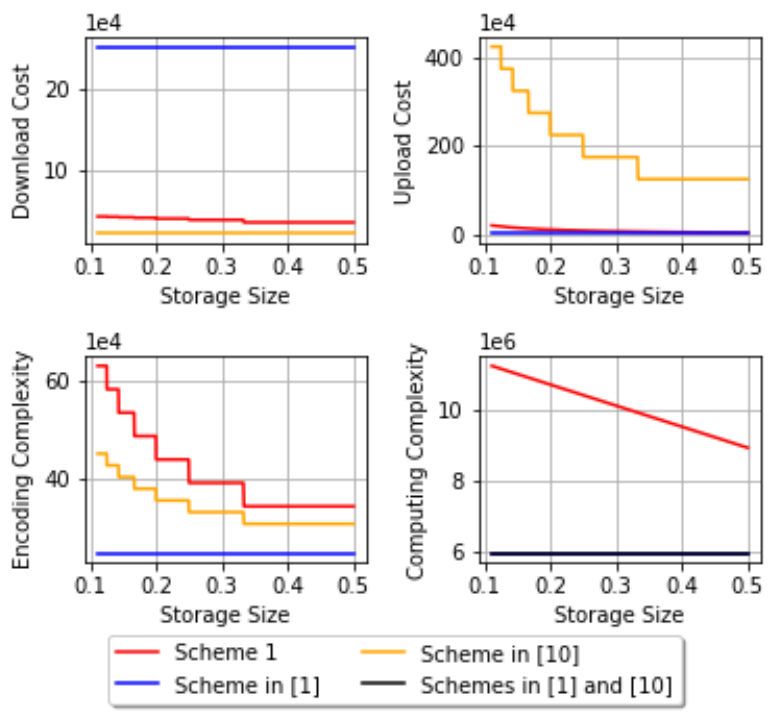}}
    \caption{Comparisons between Scheme $1$ and \cite{yang2018coded}, \cite{yangCEC} with $S = 0$ based on storage-sharing in Example \ref{ex-storage-sharing}. The blue and orange lines represent the storage-sharing of \cite{yang2018coded} and \cite{yangCEC}, respectively. The red line represents storage-sharing of Scheme $1$. The black line represents both \cite{yang2018coded} and \cite{yangCEC}.}
    \label{fig-sharing-compare}
    \end{figure}
    From Fig. \ref{fig-sharing-compare}, Scheme $1$ significantly reduces the download cost compared to the scheme in \cite{yang2018coded}, and reduces the upload cost compared to \cite{yangCEC}.
\end{example}
\subsection{Experiments on AWS EC2}
The goal is to evaluate the computation time of machines using LCSD, with cyclic assignment \cite{yang2018coded} and heterogeneous assignment \cite{wcj2021hs}, respectively. 
\subsubsection{Evaluation Setup}
We set up the system on AWS EC2 with the following configuration. 
The system consists of one t2.x2large master machine equipped with $8$ vCPUs and $32$ GiB of memory, along with $20$ worker instances. The worker 
instances include $10$ t2.large instances, each with $2$ vCPUs and $8$ GiB of memory, and $10$ t2.xlarge instances, each with $4$ vCPUs and $16$ GiB of memory, making the total number of machines $N = 20$.
The computation speeds of the $20$ instances are estimated and normalized as [$1$, $1$, $1$, $1$, $1$, $1$, $1$, $1$, $1$, $1$, $1.5$, $1.5$, $1.5$, $1.5$, $1.5$, $1.5$, $1.5$, $1.5$, $1.5$, $1.5$], respectively. 
We set $L = 5$ and vary $P \in \{0, 1, \cdots, 10\}$. For a given $P$, all available realizations in $\boldsymbol{\mathcal{N}}_{P}$ are assumed to have equal probability.
To conduct the experiments, we randomly generate a data matrix $\boldsymbol{A} \in \mathbb{F}_{1993}^{5000\times 5000}$.  In each iteration $t$, we generate randomly matrix $\boldsymbol{B}^{(t)} \in \mathbb{F}_{1993}^{5000 \times 1}$ and the available machines $\mathcal{N}_{t} \in  \boldsymbol{\mathcal{N}}_{P}$. The system then performs LCSD using Scheme $2$ with both cyclic and heterogeneous assignments, with the latter based on the estimated computation speeds.
We record the computation time of the machines during each iteration and calculate the average computation time over $5000$ iterations.

\subsubsection{Experiments Results}
\begin{figure}[H]
\centerline{\includegraphics[scale= 0.5]{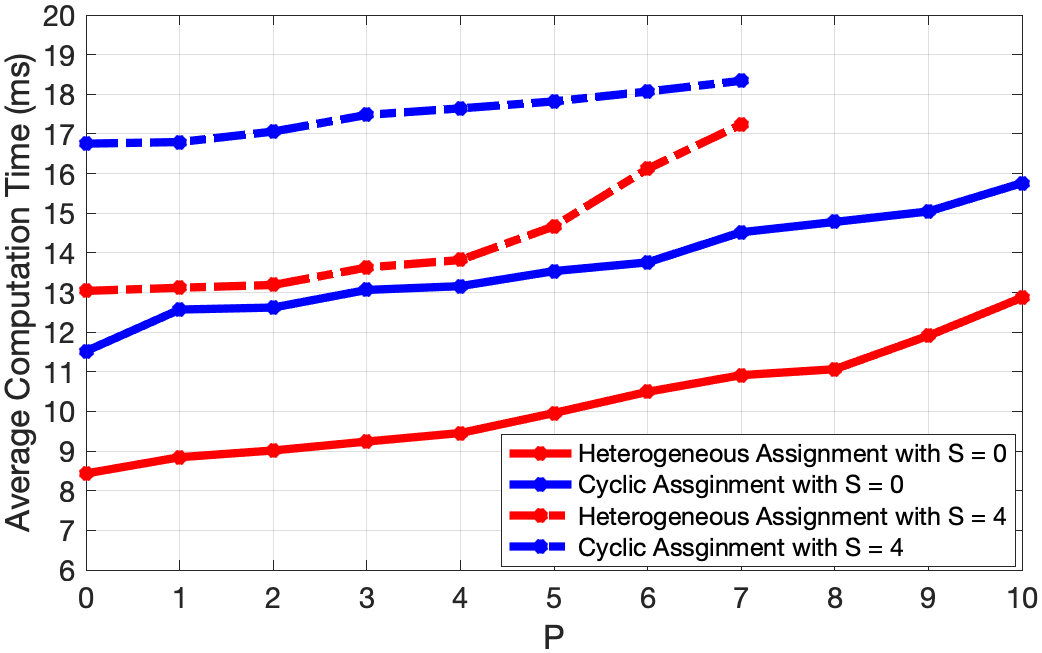}}
\caption{Experiment results when $N = 20$ and $L = 5$. The red and blue lines represent heterogeneous assignment and cyclic assignment, respectively. The solid and dash lines represent the cases of $S = 0$ and $S = 4$, respectively. \label{fig-experiment}}
\end{figure}

The experiment results are depicted in Fig. \ref{fig-experiment}.
\setlength\belowcaptionskip{-1ex}

From Fig.~\ref{fig-experiment}, we have the following observations. 
 When no straggler tolerance is considered (solid lines), heterogeneous assignment achieves a $20\%$-$30\%$ gain over cyclic assignment. 
When the system tolerates $4$ stragglers (dashed lines), the gain from heterogeneous assignment decreases from $22\%$ to $6\%$.  Notably, when $P = 7$ the average computation times of the two assignment methods are nearly identical, with only a $6\%$ gain for heterogeneous assignment. 
The reason is as follows.
In an iteration $t$ where $N_{t} = 2L+S-1 =13$, the computation load on each machine is the same for both heterogeneous and cyclic assignments. Consequently, the two assignment methods yield identical computation times. Ideally, the probability of $N_{t} = 13$ in a given iteration is ${20 \choose 7}/|\boldsymbol{\mathcal{N}}_{7}| = 56\%$. 
It means $56\%$ of the iterations would result in a zero gain from heterogeneous assignment.
However, in the experiments, with $5000$ iterations ($5000 \ll {20 \choose 7} \ll |\boldsymbol{\mathcal{N}}_{7}|$), the fraction of iterations where $|N_{t}| = 13$ is much larger than $56\%$. As a result, the majority of iterations contribute zero gain, leading to a significantly reduced overall gain of $6\%$ when $P = 7$.

\bibliographystyle{IEEEtran}
\bibliography{reference}

\begin{thebibliography}{10}
\providecommand{\url}[1]{#1}
\csname url@samestyle\endcsname
\providecommand{\newblock}{\relax}
\providecommand{\bibinfo}[2]{#2}
\providecommand{\BIBentrySTDinterwordspacing}{\spaceskip=0pt\relax}
\providecommand{\BIBentryALTinterwordstretchfactor}{4}
\providecommand{\BIBentryALTinterwordspacing}{\spaceskip=\fontdimen2\font plus
\BIBentryALTinterwordstretchfactor\fontdimen3\font minus \fontdimen4\font\relax}
\providecommand{\BIBforeignlanguage}[2]{{%
\expandafter\ifx\csname l@#1\endcsname\relax
\typeout{** WARNING: IEEEtran.bst: No hyphenation pattern has been}%
\typeout{** loaded for the language `#1'. Using the pattern for}%
\typeout{** the default language instead.}%
\else
\language=\csname l@#1\endcsname
\fi
#2}}
\providecommand{\BIBdecl}{\relax}
\BIBdecl

\bibitem{yang2018coded}
Y.~{Yang}, M.~{Interlandi}, P.~{Grover}, S.~{Kar}, S.~{Amizadeh}, and M.~{Weimer}, ``Coded elastic computing,'' in \emph{Proc IEEE ISIT}, July 2019, pp. 2654--2658.

\bibitem{KSA2021}
S.~Kiani, T.~Adikari, and S.~C. Draper, ``Hierarchical coded elastic computing,'' in \emph{Proc IEEE ICASSP}, 2021, pp. 4045--4049.

\bibitem{wcj2021hs}
N.~Woolsey, R.-R. Chen, and M.~Ji, ``Coded elastic computing on machines with heterogeneous storage and computation speed,'' \emph{IEEE Trans. on Commun.}, vol.~69, no.~5, pp. 2894--2908, 2021.

\bibitem{myjpractice}
N.~Woolsey, J.~Kliewer, R.-R. Chen, and M.~Ji, ``A practical algorithm design and evaluation for heterogeneous elastic computing with stragglers,'' in \emph{Proc IEEE GLOBECOM}, 2021, pp. 1--6.

\bibitem{DHFL2023}
S.~H. Dau, R.~Gabrys, Y.-C. Huang, C.~Feng, Q.-H. Luu, E.~J. Alzahrani, and Z.~Tari, ``Transition waste optimization for coded elastic computing,'' \emph{IEEE Trans. Inf. Theory}, vol.~69, no.~7, pp. 4442--4465, 2023.

\bibitem{usutec2022}
M.~Ji, X.~Zhang, and K.~Wan, ``A new design framework for heterogeneous uncoded storage elastic computing,'' in \emph{Proc IEEE WiOpt}, 2022, pp. 269--275.

\bibitem{XJM2023}
X.~Zhong, J.~Kliewer, and M.~Ji, ``Matrix multiplication with straggler tolerance in coded elastic computing via lagrange code,'' in \emph{Proc IEEE ICC}, 2023, pp. 136--141.

\bibitem{XJM2024}
X.~\vspace{0mm}Zhong, J.~Kliewer, and M.~Ji, ``Uncoded storage coded transmission elastic computing with straggler tolerance in heterogeneous systems,'' in \emph{IEEE ICC}, 2024, pp. 4730--4735.

\bibitem{HYWQJ2024}
W.~Huang, X.~You, K.~Wan, R.~C. Qiu, and M.~Ji, ``Decentralized uncoded storage elastic computing with heterogeneous computation speeds,'' in \emph{Proc IEEE ISIT}, 2024, pp. 1361--1366.

\bibitem{yangCEC}
Y.~Yang, M.~Interlandi, P.~Grover, S.~Kar, S.~Amizadeh, and M.~Weimer, ``Coded elastic computing,'' \emph{arXiv:1812.06411v3}, 2018.

\bibitem{XSJM2025uncodedB}
X.~Zhong, S.~Lu, J.~Kliewer, and M.~Ji, ``Uncoded download in lagrange-coded elastic computing with straggler tolerance,'' \emph{arXiv:2501.16298}, 2025.

\end{thebibliography}
\end{document}